\documentclass[reprint,twocolumn,aps,pra,amsmath,amssymb,superscriptaddress,longbibliography]{revtex4-1}

\usepackage{graphicx}
\usepackage{color}
\usepackage[colorlinks,urlcolor=cyan,citecolor=blue,linkcolor=magenta]{hyperref}
\makeatletter
\newcommand*{\rom}[1]{\expandafter\@slowromancap\romannumeral #1@}
\makeatother

\begin{document}
\title{Probing two Higgs oscillations in a one-dimensional Fermi superfluid with Raman-type spin-orbit coupling}
\author{Genwang Fan}
\affiliation{College of Physics, Qingdao University, Qingdao 266071, China}

\author{Xiao-Long Chen}
\email{xiaolongchen@swin.edu.au}
\affiliation{Department of Physics, Zhejiang Sci-Tech University, Hangzhou 310018, China}
\affiliation{Institute for Advanced Study, Tsinghua University, Beijing 100084, China}

\author{Peng Zou}
\email{phy.zoupeng@gmail.com}
\affiliation{College of Physics, Qingdao University, Qingdao 266071, China}

\date{\today}

\begin{abstract}
We theoretically investigate the Higgs oscillation in a one-dimensional Raman-type spin-orbit-coupled Fermi superfluid with the time-dependent Bogoliubov-de Gennes equations. By linearly ramping or abruptly changing the effective Zeeman field in both the Bardeen-Cooper-Schrieffer state and the topological superfluid state, we find the amplitude of the order parameter exhibits an oscillating behaviour over time with two different frequencies (i.e., two Higgs oscillations) in contrast to the single one in a conventional Fermi superfluid. The observed period of oscillations has a great agreement with the one calculated using the previous prediction [Volkov and Kogan, \href{http://jetp.ras.ru/cgi-bin/e/index/e/38/5/p1018?a=list}{J. Exp. Theor. Phys. {\bf 38}, 1018 (1974)}], where the oscillating periods are now determined by the minimums of two quasi-particle spectrum in this system. We further verify the existence of two Higgs oscillations using a periodic ramp strategy with theoretically calculated driving frequency. Our predictions would be useful for further theoretical and experimental studies of these Higgs oscillations in spin-orbit-coupled systems.
\end{abstract}
 
\maketitle


\section{INTRODUCTION}

Collective excitation is important dynamics of many-body quantum system, and becomes an interesting research topic in all realm of physics. As a kind of gapped collective excitation, the Higgs mode is a quantum phenomenon investigated in superconductors~\cite{littlewoodPRL1981,littlewoodPRB1982,sooryakumarPRL1980,sooryakumarPRB1981}, magnetic materials~\cite{matsumotoPRB2004,rueggPRL2008}, and ultracold atoms in continuous or lattice system~\cite{scottPRA2012,altmanPRL2002,polletPRL2012,bissbortPRL2011,endresNAT2012}. A review paper about Higgs mode in condensed matter physics can be found in Ref.~\cite{pekkerCMP2015}. Physically the Higgs mode is described by the amplitude fluctuation of the order parameter, which is different from the gapless Goldstone excitation related to the phase fluctuation of order parameter.

While the appearance of Goldstone mode is easy to observe when continuous symmetries are broken, stable Higgs modes require additional symmetry to stop them from rapidly decaying into other low-energy excitations. In high-energy physics, the stability of Higgs mode is ensured by Lorentz invariance, whose role is replaced by particle-hole symmetry in condensed matter physics. The famous Bardeen-Cooper-Schrieffer (BCS) Hamiltonian describing a weakly interacting superconductor is a typical example of hosting a stable Higgs mode with particle-hole mechanism~\cite{littlewoodPRL1981,littlewoodPRB1982}, and related evidence has also been found in conventional BCS superconductors~\cite{sooryakumarPRL1980,matsunagaPRL2013,shermanNP2015}. The same BCS theory, which is usually called Bogoliubov-de Gennes (BdG) mean field theory, is also widely used to study the ultracold Fermi gases. The Higgs mode has also been theoretically investigated in the BCS-Bose Einstein Condensate (BEC) crossover of Fermi superfluid~\cite{yuzbashyanPRL2006,scottPRA2012,hannibalPRA2015} with a time-dependent Bogoliubov-de Gennes (BdG) simulation. The order parameter has a close connection with the condensate fraction~\cite{altmanPRL2005,peraliPRL2005,matyjaskiewiczPRL2008}. Following this relation, experimentally the Higgs mode has been observed in a strongly interacting fermionic superfluid with radiofrequency field technique~\cite{behrleNP2018}.

To excite the Higgs mode in ultracold Fermi gases, generally one can resort to the modulation of all parameters which can decide the order parameter, e.g., the interaction parameter $1/\left(k_{F}a\right)$ in the BCS-BEC crossover~\cite{scottPRA2012,liuPRA2016,hanPRA2016,kurkjianPRL2019}. In 1974, Volkov and Kogan studied the response of superconductors in the presence of a small perturbation with the Green's functions,  and found that the order parameter $\left|\Delta\right|$ oscillates with the period $\pi\hbar/\Delta_{\rm{gap}}$, where the $\Delta_{\rm{gap}}$ is the energy gap in the spectrum of fermionic excitations~\cite{volkovSPJ1974}. This also indicates that the Higgs mode is greatly influenced by the single-particle excitation. Usually the Higgs mode is mixed and coupled with the continuum spectrum of the single-particle excitation in many Fermi superfluids and thus we will call it a Higgs oscillation instead in the remaining text. Since the development of artificial gauge field in Fermi superfluid~\cite{wangPRL2012,cheukPRL2012}, more control knobs, like effective Zeeman field and spin-orbit coupling strength, can be brought in to perturb the amplitude of order parameter. Higgs oscillation is expected to display richer and much interesting dynamical behavior in spin-orbit coupled (SOC) degenerate Fermi gases. Previously topological phase transition of quench dynamics and dynamical phases had been studied in SOC Fermi superfluid~\cite{wangNJP2015,dongNC2015}. But to date there are quite few discussions to introduce the Higgs oscillation and its physical properties in SOC Fermi superfluid. In this paper, we will introduce two kinds of Higgs oscillations with different periods in SOC Fermi superfluid.

In this work, motivated by previous theoretical studies and recent experiments, we explore the fascinating Higgs oscillation in a one-dimensional (1D) SOC Fermi superfluid and aim to characterize two distinct Higgs oscillations by studying the related dynamic behaviour. With the help of time-dependent BdG equation, we first investigate the properties of the order parameter as well as the excitation spectrum on the tunable effective Zeeman field in different phase regimes. By introducing three time-dependent ways to tune the effective Zeeman field, we then obtain the oscillating behaviours of the amplitude of the order parameter in both the BCS and topological phases. Finally, by means of a Fourier analysis, we numerically calculate the oscillating frequency or period to straightforwardly characterize the Higgs oscillation, and compare it with the previous theoretical prediction of Volkov and Kogan in both two phases.

The paper is organized as follows. In the next section, we will briefly introduce the model and Hamiltonian of a SOC Fermi superfluid with the mean-field theory. In Sec.~\ref{sec:results}, we probe and investigate two Higgs oscillations in both the BCS and topological states, by tuning the effective Zeeman field in three different ways and investigating the oscillating behaviors of the amplitude of the order parameter. Finally, we summarize and draw conclusions in Sec.~\ref{sec:conclusion}.

\section{MODEL AND HAMILTONIAN}

We consider a 1D superfluid Fermi gas with Raman-type spin-orbit coupling effect. The system can be described by a single-channel model Hamiltonian $H=\int dx\left[\mathcal{H}_{0}+\mathcal{H}_{int}\right]$, where
\begin{equation}
\mathcal{H}_{0}=\begin{bmatrix}
\psi_{\uparrow}^{\dagger}\left(x\right),\psi_{\downarrow}^{\dagger}\left(x\right)
\end{bmatrix}
\left(\mathcal{H}_{s}+\lambda\hat{k}_{x}\sigma_{y}-h\sigma_{z}\right)
\begin{bmatrix}
\psi_{\uparrow}\left(x\right)\\
\psi_{\downarrow}\left(x\right)
\end{bmatrix}
\end{equation}
is the SOC single-particle part in a uniform system and 
\begin{equation}
\mathcal{H}_{int}=g_{_\mathrm{1D}}\psi_{\uparrow}^{\dagger}\left(x\right)\psi_{\downarrow}^{\dagger}\left(x\right)\psi_{\downarrow}\left(x\right)\psi_{\uparrow}\left(x\right)
\end{equation}
is the interaction Hamiltonian with $g_{_\mathrm{1D}}=-\gamma\hbar^{2}n_{0}/m$ describing an attractive $s$-wave contact interaction between two spin states $\left(\sigma=\uparrow,\downarrow\right)$. $n_{0}$ is the bulk density, and $\gamma$ denotes a dimensionless interaction parameter. Here $\psi_{\sigma}$ or $\psi_{\sigma}^{\dagger}$ is the field operator that annihilates or creates a spin $\sigma$ atom with mass $m$. $\mathcal{H}_{s}=-\hbar^{2}\partial_{x}^{2}/\left(2m\right)-\mu$ describes the motion of free atoms  with chemical potential $\mu$. The term $\lambda\hat{k}_{x}\sigma_{y}-h\sigma_{z}$, with the momentum operator $\hat{k}_{x}=-i\partial/\partial_{x}$ and Pauli matrices $\sigma_{y}$ and $\sigma_{z}$, is induced by the Raman process, describing a synthetic spin-orbit coupling with a strength $\lambda\equiv\hbar^{2}k_{R}/m$ and an effective Zeeman field $h=\Omega_{R}/2$. Here $k_{R}$ and $\Omega_{R}$ are the recoil momentum and the Rabi frequency of Raman laser beams, respectively. In the following, we will set $\hbar=1$ for simplicity.

We use the standard mean-field theory to solve the single-channel model Hamiltonian. Within the mean-field approximation, we define an order parameter $\Delta\left(x\right)\equiv-g_{_\mathrm{1D}}\left\langle \psi_{\downarrow}\left(x\right)\psi_{\uparrow}\left(x\right)\right\rangle $, and the interaction Hamiltonian is decoupled as
\begin{equation}
\mathcal{H}_{int}^{MF}\simeq-\left(\Delta\psi_{\uparrow}^{\dagger}\psi_{\downarrow}^{\dagger}+\Delta^{*}\psi_{\downarrow}\psi_{\uparrow}\right)-\left|\Delta\right|^{2}/g_{_\mathrm{1D}}.
\end{equation}
After the standard Bogoliubov transformation $\psi_{\sigma}=\sum_{\eta}\left[u_{\sigma\eta}e^{-iE_{\eta}t}c_{\eta}+v_{\sigma\eta}^{*}e^{iE_{\eta}t}c_{\eta}^{\dagger}\right]$ to all field operators of mean-field Hamiltonian, we obtain the BdG equations
\begin{equation} \label{eq:bdg-equation}
H_{BdG}\phi_{\eta}\left(x\right)=E_{\eta}\phi_{\eta}\left(x\right)
\end{equation}
 in a Nambu spinor representation with BdG Hamiltonian
\begin{equation}
H_{BdG}\equiv\left[\begin{array}{cc}
\begin{array}{cc}
\mathcal{H}_{S}-h & -\lambda\partial/\partial x\\
\lambda\partial/\partial x & \mathcal{H}_{S}+h
\end{array} & \begin{array}{cc}
0 & -\Delta\left(x\right)\\
\Delta\left(x\right) & 0
\end{array}\\
\begin{array}{cc}
0 & \Delta^{*}\left(x\right)\\
-\Delta^{*}\left(x\right) & 0
\end{array} & \begin{array}{cc}
-\mathcal{H}_{S}+h & \lambda\partial/\partial x\\
-\lambda\partial/\partial x & -\mathcal{H}_{S}-h
\end{array}
\end{array}\right],
\end{equation}
the quasi-particle wave function $\phi_{\eta}=[u_{\uparrow\eta},u_{\downarrow\eta},\upsilon_{\uparrow\eta},\upsilon_{\downarrow\eta}]^{T}$ and the corresponding quasi-particle eigenenergy $E_{\eta}$. The BdG equations above should be solved self-consistently with the order parameter equation
\begin{equation}
\Delta(x)=-\frac{g_{_\mathrm{1D}}}{2}\underset{\eta}{\sum}\left[u_{\uparrow\eta}\upsilon_{\downarrow\eta}^{*}f(E_{\eta})+u_{\downarrow\eta}\upsilon_{\uparrow\eta}^{*}f(-E_{\eta})\right]\label{eq:del}
\end{equation}
and the density equation
\begin{figure}[t]
\includegraphics[width=0.48\textwidth]{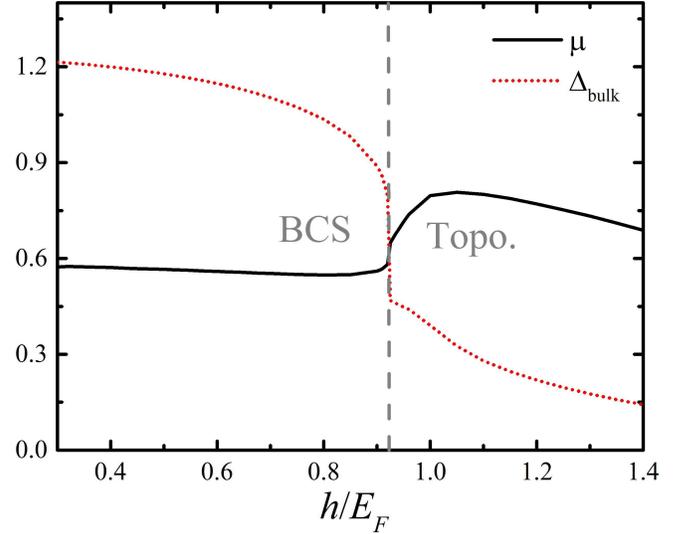}
\caption{\label{fig:fig1} The chemical potential $\mu$ and the bulk value of order parameter $\Delta_\mathrm{bulk}$ as a function of the effective Zeeman field $h$ in a 1D SOC Fermi gas with a box potential. The vertical dashed line indicates the position of the phase transition between the BCS and topological phases at $h_{c}\simeq0.92E_{F}$.}
\end{figure}
\begin{equation}
n(x)=\frac{1}{2}\underset{\sigma\eta}{\sum}\left[|u_{\sigma\eta}|^{2}f(E_{\eta})+|\upsilon_{\sigma\eta}|^{2}f(-E_{\eta})\right], \label{eq:den}
\end{equation}
where $f(E)=1/[e^{E/k_{B}T}+1]$ is the Fermi-Dirac distribution function at a temperature $T$. It is important to note that the use of Nambu spinor representation doubles the size of Hilbert space of the system. As a result, there is a particle-hole symmetry in the Bogoliubov solutions: for any ``particle'' solution with the wave function $\phi_{\eta}^{\left(p\right)}=\left[u_{\uparrow\eta},u_{\downarrow\eta},\upsilon_{\uparrow\eta},\upsilon_{\downarrow\eta}\right]^{T}$ and energy $E_{\eta}^{\left(p\right)}\geq0$, we can always find the other ``hole'' solution with a wave function $\phi_{\eta}^{\left(h\right)}=\left[\upsilon_{\uparrow\eta}^{*},\upsilon_{\downarrow\eta}^{*},u_{\uparrow\eta}^{*},u_{\downarrow\eta}^{*}\right]{}^{T}$ and energy $E_{\eta}^{\left(h\right)}=-E_{\eta}^{\left(p\right)}\leq0$. Generally these two states describe the same physical state. To remove this redundancy, we have added an extra factor of $1/2$ in the expressions for the order parameter \eqref{eq:del} and total density \eqref{eq:den}. In the following discussions, we focus at zero temperature with a typical interaction strength $\gamma=\pi$, the SOC strength $\lambda=1.5E_{F}/k_{F}$. As shown in Fig. \ref{fig:fig1}, the system undergoes a phase transition from a BCS superfluid to a topological superfluid when continuously increasing the effective Zeeman field $h$ over a critical value $h_{c}\simeq0.92E_{F}$~\cite{kongPRA2021}, where the chemical potential $\mu$ and the bulk order parameter $\Delta_\mathrm{bulk}$ present a jump change. We need to emphasize that the value of $h_{c}$ will be slightly influenced by some parameters in the numerical calculation such as the size of box and the energy cutoff.

\begin{figure}
\includegraphics[width=0.48\textwidth]{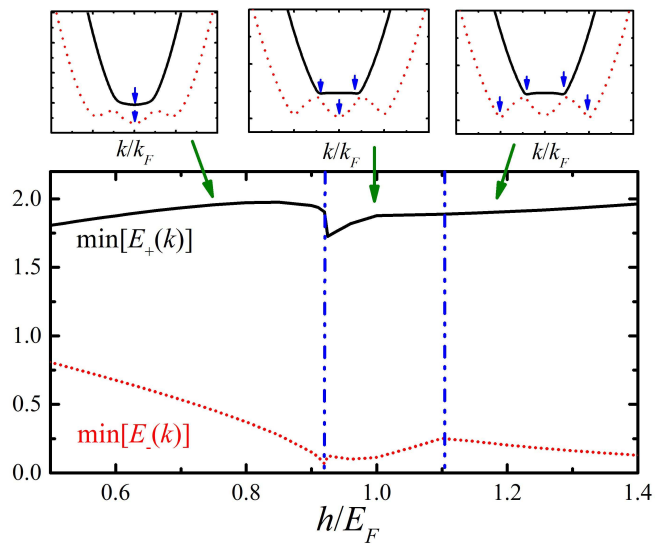}
\caption{\label{fig:fig2} The positions and values of minimum in two quasi-particle energy branches $E_{\pm}\left(k\right)$ as a function of the effective Zeeman field $h$.}
\end{figure}

In a uniform and infinite system, the continuous momentum $k$ is a good quantum number. Thus, it is possible to get an analytic expression of four quasi-particle eigenenergy $E_{\pm}^{\left(p\right)}\left(k\right)=-E_{\pm}^{\left(h\right)}\left(k\right)\equiv E_{\pm}\left(k\right)$ in Eq.~\eqref{eq:bdg-equation} with~\cite{weiPRA2012}
\begin{equation}
E_{\pm}\left(k\right)=\sqrt{E_{k}^{2}+h^{2}+k^{2}\lambda^{2}\pm2\sqrt{\xi_{k}^{2}k^{2}\lambda^{2}+E_{k}^{2}h^{2}}},\label{eq:energy}
\end{equation}
where $\xi_{k}=k^{2}/\left(2m\right)-\mu$ and $E_{k}=\sqrt{\xi_{k}^{2}+\Delta^{2}}$. The excitation of the Higgs oscillation is closely related to the minimum of quasi-particle energy~\cite{volkovSPJ1974}. In Fig. \ref{fig:fig2}, we present the positions and values of the minimum in two positive quasi-particle energy branches $E_{\pm}\left(k\right)$ as a function of the Zeeman field $h$. Typically there are three regimes separated by $h_{c}\approx0.92E_{F}$ and $h_{sp}\approx1.1E_{F}$, where the locations of minimum in two energy branches marked by blue arrows in three upper panels are different. The locations of minimum are both at $k=0$ when $h<h_{c}$, then the minimum of $E_{+}(k)$ is shifted to a nonzero momentum while the one in $E_{-}(k)$ is still at $k=0$ when $h_{c}<h<h_{sp}$, and finally both of them are shifted to a nonzero momentum when $h>h_{sp}$.

\section{RESULTS AND DISCUSSION} \label{sec:results}

In this work, a 1D SOC Fermi superfluid is taken into account where the quasi-1D geometry is usually realized by applying a strong confinement along both $y$ and $z$ axes in a three-dimensional (3D) system~\cite{cazalillaRMP2011,guanRMP2013}. In general, the order parameter is determined by the realistic parameters of the system, such as the interaction strength $\gamma$ , the SOC strength $\lambda$ and the effective Zeeman field $h$. The interaction strength can be well controlled by both confinement and Feshbach resonances as discussed in references~\cite{olshaniiPRL1998,hallerSCI2009,hallerPRL2010,pengPRA2010,pengPRA2011,pengPRL2014}. However, it is tough to change the interaction strength rapidly or in a very short time scale. In addition, the SOC strength $\lambda$ can not be tuned over a large range in ultracold atoms experiments. Thus, we choose the effective Zeeman field $h$ as the ramping parameter in this work. First, the Zeeman field determines directly the topological structure of the ground state as shown in Fig.~\ref{fig:fig1} and we can discuss for different cases. Besides, the effective Zeeman field can be feasibly tuned in a long or short time scale by the laser intensity or the detuning in the experiments of the SOC Fermi gases. These experiment features have been introduced in Refs.~\cite{linNat2011,wangPRL2012,cheukPRL2012,zhangPRL2012,williamsPRL2013,quPRA2013}. In order to excite and investigate the Higgs oscillation, we begin with an initial Zeeman field $h_{i}$ to calculate self-consistently its ground state, and then vary $h$ over time in a slow linear ramp way or by abruptly changing to reach a final Zeeman field $h_{f}$. Therefore, the dynamics of the order parameter can be then studied by solving the time-dependent BdG equation
\begin{equation}
i\frac{\partial}{\partial t}\phi_{\eta}\left(x,t\right)=H_{BdG}\phi_{\eta}\left(x,t\right).
\end{equation}
\begin{figure}
\includegraphics[width=0.48\textwidth]{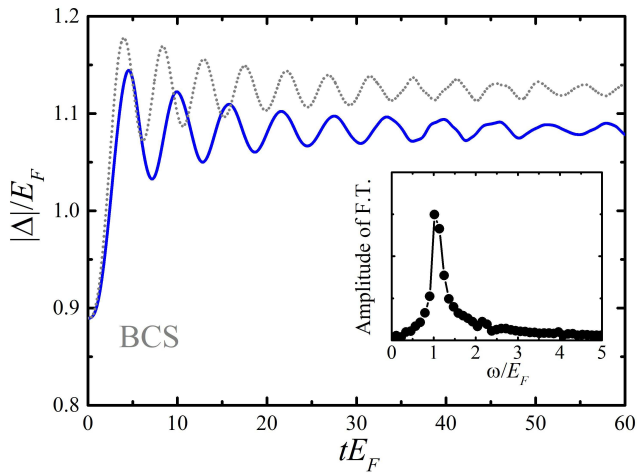}
\caption{\label{fig:fig3}The oscillation of the order parameter $\left|\Delta(t)\right|$ during a slow ramp in the BCS state. The Zeeman field is tuned adiabatically from $h_{i}=0.9E_{F}$ to $h_{f}=0.7E_{F}$ (blue solid line), and from $h_{i}=0.9E_{F}$ to $h_{f}=0.6E_{F}$ (gray dotted line). The inset is the Fourier analysis of the blue solid line with a peak at $\omega\simeq1.02E_{F}$ indicating the oscillating frequency.}
\end{figure}

\subsection{Slow ramp of the effective Zeeman field}

We begin with a slow linear ramp of the Zeeman field $h$, namely $h(t)=h_{i}+\left(h_{f}-h_{i}\right)t/t_{0}$, in which $t_{0}$ is the time consumed to arrive at the final Zeeman field $h_{f}$. Generally the order of $t_{0}$ can not be very small to make the system evolve in an almost adiabatic process. So $t_{0}$ should be at least in an order of $1/E_{F}$.

We first choose an initial Zeeman field $h_{i}=0.9E_{F}$, and linearly decrease $h$ to $h_{f}=0.7E_{F}$ in a time regime $t_{0}=4/E_{F}$. Obviously $h_{i},h_{f}<h_{c}$, which means that the system is in the BCS state. As shown by a smooth blue solid line in Fig. \ref{fig:fig3}, the amplitude of the order parameter first increases monotonically from $0.89E_{F}$ (i.e., the equilibrium value of $|\Delta_{eq}|$ at $h=h_{i}$ obtained from Eq.~\eqref{eq:bdg-equation}), and then oscillates around an average value $\Delta_{\infty}=1.08E_{F}$ (i.e., almost the equilibrium value of $|\Delta_{eq}|$ at $h=h_{f}$). The oscillation period can be determined by the Fourier analysis of the oscillation of the order parameter. As shown in the inset, there's 
a frequency peak at $\omega\simeq1.02E_{F}$ in the Fourier analysis, giving rise to an oscillation period at about $T_{1}=2\pi/\omega=6.16/E_{F}$. In addition, Volkov and Kogan predicted that the oscillation period of the Higgs oscillation should be~\cite{volkovSPJ1974}
\begin{equation}  \label{eq:Tana}
T_\mathrm{VK}=\pi/\Delta_\mathrm{gap},
\end{equation}
where $\Delta_\mathrm{gap}$ is usually the minimum of quasi-particle energy, and here
\begin{equation}
\Delta_\mathrm{gap}=\Delta_\mathrm{gap}^{\left(-\right)}=\mathrm{min}\left[E_{-}\left(k\right)\right]
\end{equation}
is equal to a half of the minimum energy to break a Cooper pair. Here the chemical potential $\mu$ and the Zeeman field $h$ in Eq.~\eqref{eq:energy} should use their values at the final state ($h=h_{f}$), while the order parameter should take the value of $\Delta_{\infty}$~\cite{scottPRA2012}. And we find $T_\mathrm{VK}^{\left(-\right)}=6.09/E_{F}$, which is quite close to the numerical value $T_{1}\simeq6.16/E_{F}$ obtained from the Fourier analysis with a deviation rate about $1\%$. For comparison, we also simulate with another set of parameters (i.e., from $h_{i}=0.9E_{F}$ to $h_{f}=0.6E_{F}$ denoted by gray dotted line), and the difference rate between $T_\mathrm{VK}$ and the numerically calculated one is also around $1\%$. So the Higgs oscillation here is closely related to the excitation in the lower branch of quasi-particle spectrum, and we call it the low Higgs oscillation in the following discussion.

\begin{figure}
\includegraphics[width=0.48\textwidth]{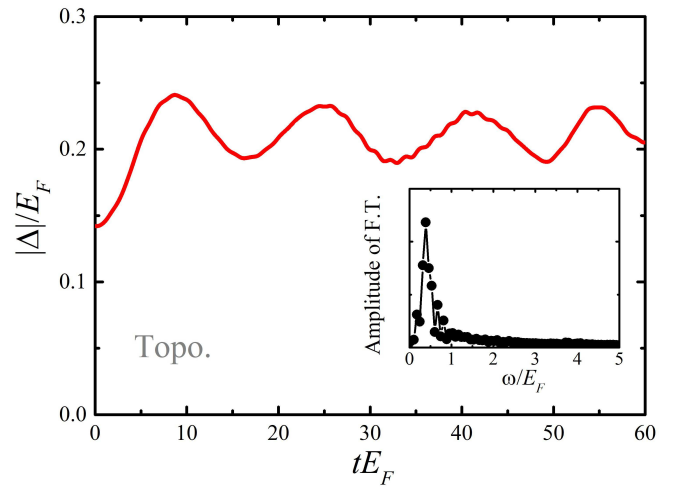}
\caption{\label{fig:fig4} The oscillation of the order parameter $\left|\Delta(t)\right|$ during a slow ramp in the topological state. The Zeeman field is tuned from $h_{i}=1.4E_{F}$ to $h_{f}=1.2E_{F}$. The Fourier analysis in the inset reveals an oscillating frequency at $\omega\simeq0.39E_{F}$.}
\end{figure}

We now turn to consider the topological superfluid where both values of the initial and final Zeeman field are larger than the critical one $h_{c}$. The Zeeman field is linearly tuned from $h_{i}=1.4E_{F}$ to $h_{f}=1.2E_{F}$. We can know from Fig.~\ref{fig:fig1} that increasing the Zeeman field in the topological state will generally decreases the corresponding order parameter in equilibrium. In Fig. \ref{fig:fig4}, we find a similar oscillation in the amplitude of the order parameter as in the case of the BCS superfluid, around an almost equilibrium value $\Delta_{\infty}=0.21E_{F}$. Similarly, we can figure out the oscillation period $T_{1}\simeq16.11/E_{F}$ from the Fourier analysis, which also agrees well with the theoretical prediction $T_\mathrm{VK}^{\left(-\right)}=16.01/E_{F}$ within a $1\%$ deviation. Obviously this is also a low Higgs oscillation, originated from the excitation in the lower quasi-particle spectrum $E_{-}(k)$. It should also be noted that there are some tiny sawtooth-like structures in the oscillation curve at relatively large time, which make the curve not so smooth. In fact these detailed structures are closely associated to the other Higgs oscillation which will be discussed in the next subsection.

Overall, we find that the oscillation period of the Higgs oscillation obtained numerically from its dynamics agrees well with Volkov and Kogan's prediction in both the BCS superfluid and the topological one. Moreover, we also run simulation and make the Zeeman field $h$ come into the regime $h_{c}<h<h_{sp}$ , and investigate the Higgs oscillation there. However, we find a complex oscillating behaviour in the order parameter, and the numerical result of the period is quite far away from $T_\mathrm{VK}$ due to the rapid variation of $\mu$ and $\Delta_\mathrm{bulk}$ (see Fig.~\ref{fig:fig1}), or the switch of the position of minimum in the spectrum $E_{\pm}\left(k\right)$ (see Fig.~\ref{fig:fig2}). Here we argue that these two reasons make Volkov and Kogan's prediction can not work well in this regime.
\begin{figure}[t]
\includegraphics[width=0.48\textwidth]{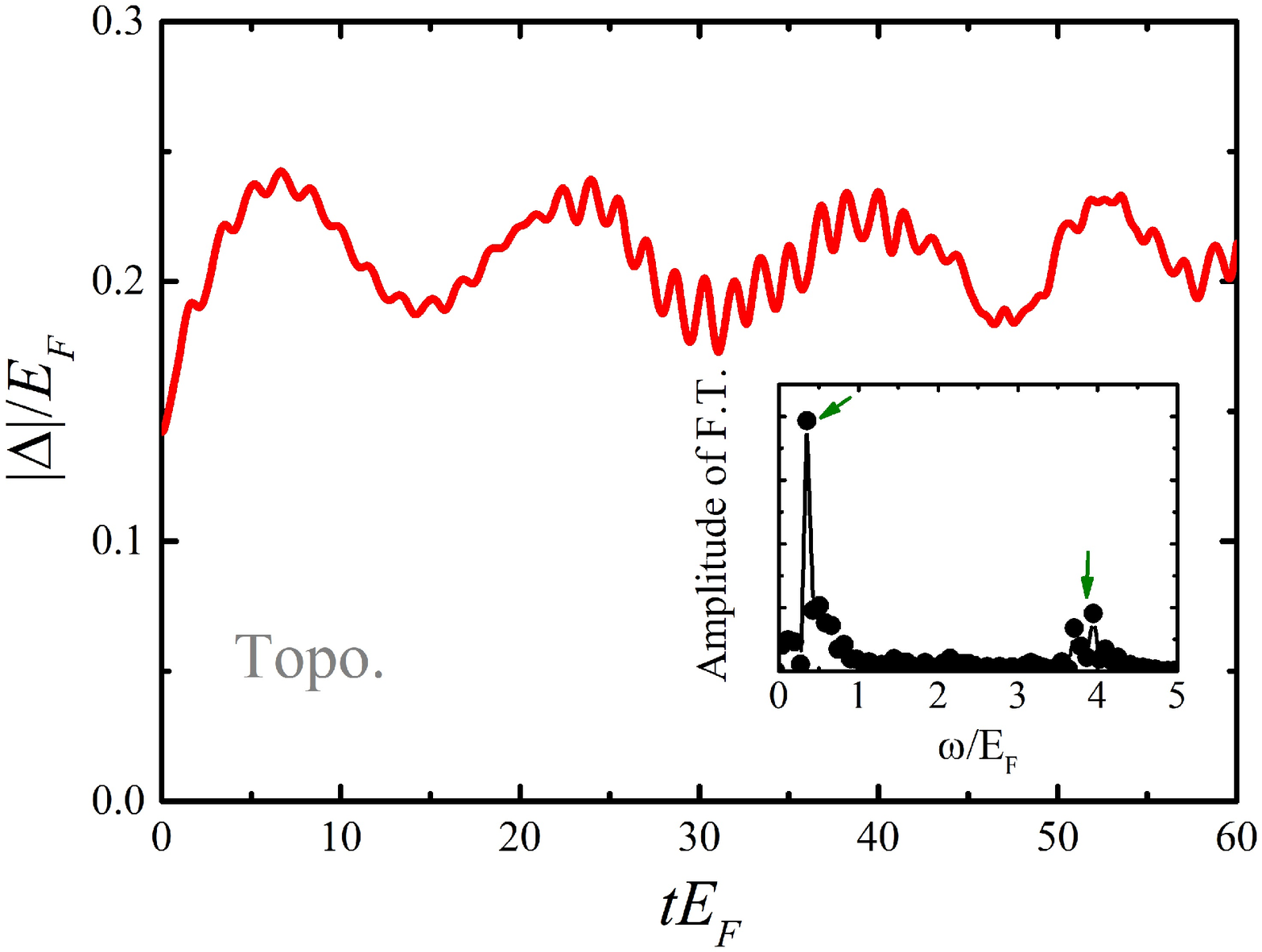}
\caption{\label{fig:fig5}The oscillation of the order parameter $\left|\Delta(t)\right|$ after an abrupt ramp in the topological state. The Zeeman field is changed suddenly from $h_{i}=1.4E_{F}$ to $h_{f}=1.2E_{F}$ at time $t=0$. The Fourier analysis in the inset reveals two oscillating frequencies at $\omega\simeq0.35E_{F}$ and $3.91E_{F}$.}
\end{figure}

\subsection{Abrupt ramp of the effective Zeeman field}

In this subsection, we consider an abrupt way to vary the Zeeman field, and investigate the following quench dynamics of the system. Similarly we prepare the system in the ground state at an initial Zeeman field $h_{i}$, and then change immediately the Zeeman field to its destination value $h_{f}$ at time $t=0$.

We first discuss the case in the topological superfluid by suddenly tuning the Zeeman field from $h_{i}=1.4E_{F}$ to $h_{f}=1.2E_{F}$ and study the quench dynamics of the system. The result of the oscillating behaviour in the order parameter and the associated Fourier analysis are present in Fig. \ref{fig:fig5}. In contrast to the single oscillation period in the conventional 3D Fermi gas~\cite{scottPRA2012}, we find that there exist two distinct periods in the amplitude of the order parameter oscillating around $\Delta_{\infty}=0.21E_{F}$. $\Delta_{\infty}$ is usually smaller than the equilibrium value of the order parameter at $h=h_f$ in the case of an abrupt ramp. The bigger period originates from the excitation in the lower branch of the energy spectrum, as we discussed in the last subsection and in Fig. \ref{fig:fig4}. However, we find that the smaller period, i.e., the sawtooth-like structure in the oscillation, can be well explained by the excitation to the higher branch $E_{+}\left(k\right)$ in Eq.~\eqref{eq:energy} in the quasi-particle energy spectrum, giving rise to the other excitation gap energy
\begin{equation}
\Delta_\mathrm{gap}^{\left(+\right)}=\mathrm{min}\left[E_{+}\left(k\right)\right]
\end{equation}
for calculating the theoretical value using Volkov and Kogan's prediction~\cite{volkovSPJ1974}. The existence of two types of the Higgs oscillation can be also seen clearly from the Fourier analysis in the inset of Fig.~\ref{fig:fig5} which presents two frequency peaks at $\omega\simeq0.35E_{F}$ and $\omega\simeq3.91E_{F}$ marked by two arrows. The low-frequency peak represents the low Higgs oscillation coming from the lower energy branch $E_{-}\left(k\right)$, while the high-frequency peak supports the other Higgs oscillation with a smaller period $T_{2}\simeq1.60/E_{F}$. This Higgs oscillation with a small period can be called the high Higgs oscillation, and has an about $4\%$ deviation from Volkov and Kogan's prediction $T_\mathrm{VK}^{\left(+\right)}=\pi/\Delta_\mathrm{gap}^{\left(+\right)}=1.66/E_{F}$ using the higher branch in the quasi-particle spectrum. Here, the ratio between two periods is about $T_{1}/T_{2}\approx10$, sufficiently large to make these two Higgs oscillations can be clearly distinguished.
\begin{figure}[t]
\includegraphics[width=0.48\textwidth]{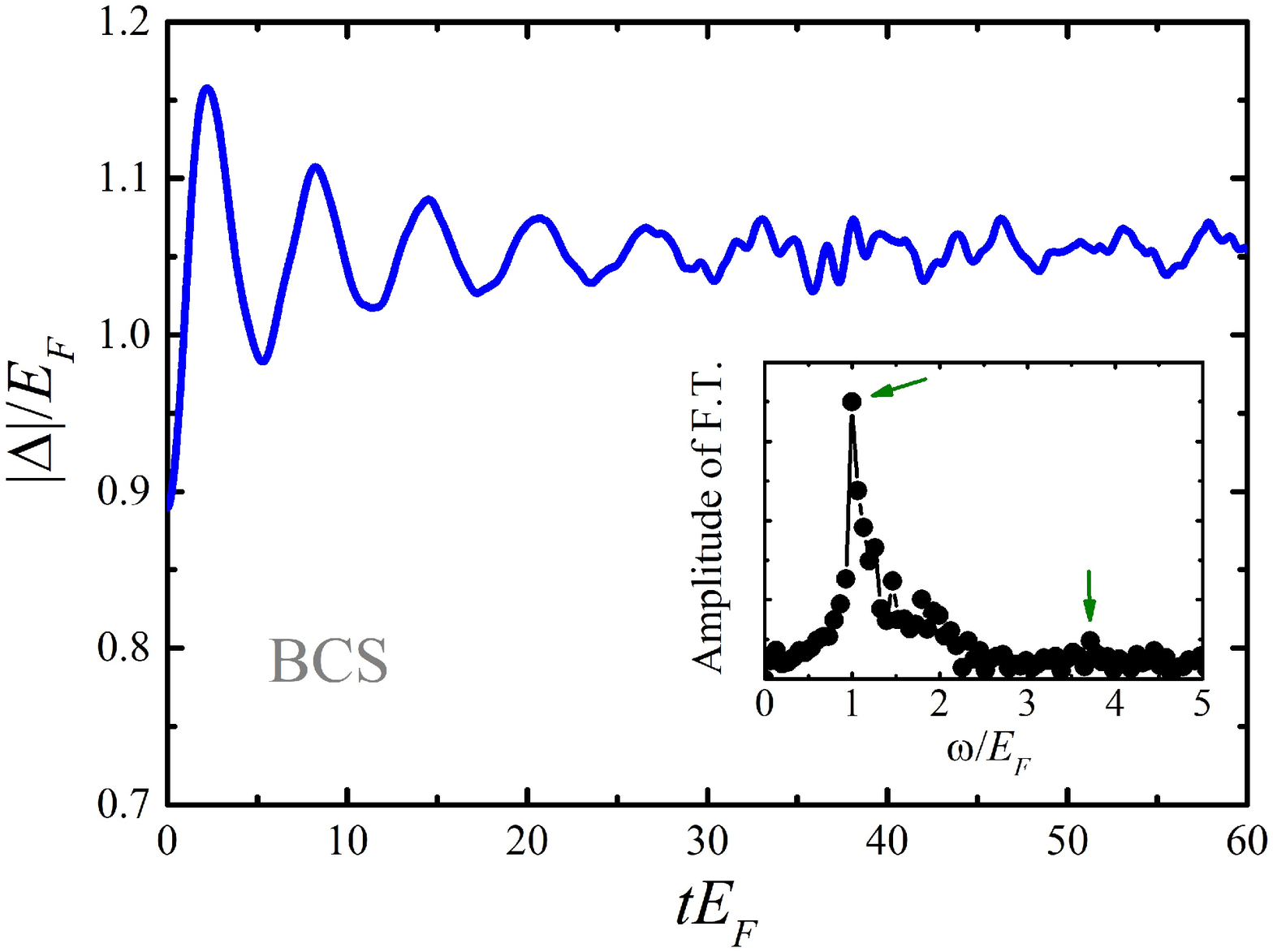}
\caption{\label{fig:fig6}The oscillation of the order parameter $\left|\Delta(t)\right|$ after an abrupt ramp in the BCS state. The Zeeman field is changed suddenly from $h_{i}=0.9E_{F}$ to $h_{f}=0.7E_{F}$ at time $t=0$. The Fourier analysis in the inset reveals two oscillating frequencies at $\omega\simeq1.00E_{F}$ and $3.72E_{F}$.}
\end{figure}

Likewise, we then turn to consider the existence of high Higgs oscillation in the quench dynamics of the BCS state, which has not been probed in the case of a slow ramp as in Fig.~\ref{fig:fig3}. We illustrate the results in Fig.~\ref{fig:fig6} by preparing a ground state at $h_{i}=0.9E_{F}$ and then suddenly changing the Zeeman field to $h_{f}=0.7E_{F}$ at time $t=0$. In general, the order parameter $|\Delta|$ oscillates around $\Delta_{\infty}=1.05E_{F}$, and displays an almost clear period for $t<25/E_{F}$. However, a complex oscillation behaviour turns out at larger time, and makes it very tough to distinguish the periodic oscillation by naked eyes. Similarly, by means of the Fourier analysis of the oscillation dynamics, we can also find two frequency peaks marked by two arrows in the inset of Fig. \ref{fig:fig6}. Using the peak frequency, the calculated periods of these two periodic oscillations are $T_{1}=6.31E_{F}$ and $T_{2}=1.69E_{F}$ respectively, which just fit well with Volkov and Kogan's prediction $T_\mathrm{VK}^{\left(-\right)}=6.41/E_{F}$ and $T_\mathrm{VK}^{\left(+\right)}=1.66/E_{F}$ using two energy branches in the quasi-particle spectrum. Compared with the topological case in Fig. \ref{fig:fig5}, the low-frequency peak of the low Higgs oscillation here is still remarkable, while the signal of the high Higgs oscillation is relatively much weaker. In fact similar to other resonance phenomena, these two Higgs oscillations are always coupled with each other, only a large period (or frequency) contrast can help to distinguish them. However, the period ratio $T_{1}/T_{2}\approx3.73$ here is much smaller than the one in the topological case, which is consistent with the expectation from Fig. \ref{fig:fig2}, i.e., the minima of two energy branches approaching each other. Thus, these factors make two Higgs oscillations tangled with each other and display a complex dynamical behaviour in Fig. \ref{fig:fig6}.
\begin{figure}[t]
\includegraphics[width=0.48\textwidth]{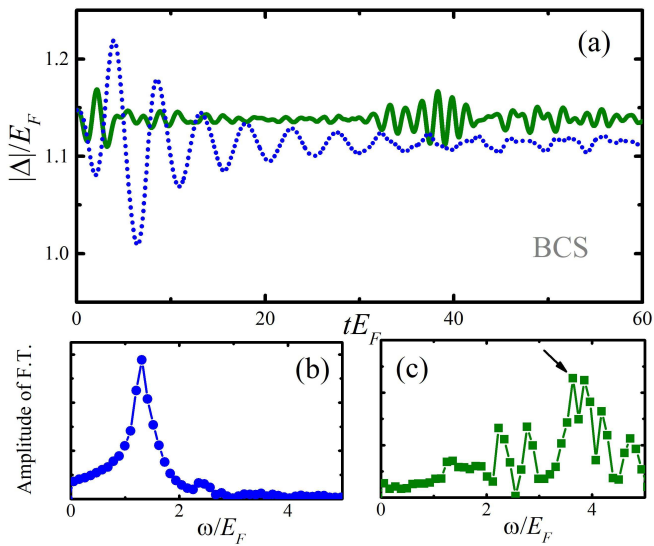}
\caption{\label{fig:fig7}(a) The oscillation of the order parameter $\left|\Delta(t)\right|$ after a periodic ramp in Eq.~\eqref{eq:hp} at $h_{i}=0.6E_{F}$. (b) and (c) are the corresponding Fourier analysis. Solid and dotted lines are the high and low Higgs oscillations, respectively.}
\end{figure}

\subsection{Further verification of two Higgs oscillations}

To further probe and study these two Higgs oscillations in both the BCS and topological states, we introduce a new way to tune the Zeeman field following the resonance theory. In the last two sections, we find that the Volkov and Kogan's prediction $T_\mathrm{VK}^{\left(-\right)}$ and $T_\mathrm{VK}^{\left(+\right)}$ agree well with the corresponding numerical results. Thus, we can use these frequencies calculated theoretically as a driving frequency to excite two Higgs oscillations respectively in only $1.5$ oscillation periods (longer driving-resonance time can do help to strengthen the resonance effect and is beneficial to the associated Fourier analysis), and stop driving in the following time, namely
\begin{equation}\label{eq:hp}
h(t)=h_{i}+A\sin{\left(2\pi t/T_\mathrm{VK}^{(\pm)}\right)},
\end{equation}
with $A$ being a small amplitude of the perturbation. In fact a larger value of $A$ will not only strengthen the amplitude of oscillation, but also possibly make the system comes into different regimes as shown in Fig.~\ref{fig:fig2}. So we choose $A=0.1E_F$ in the following discussions. With this periodic ramp strategy described above, we can then investigate its evolving dynamics at an initial Zeeman field $h_{i}$.

In the BCS state with a Zeeman field $h_{i}=0.6E_{F}$, we find two different oscillations in the amplitude of the order parameter as anticipated. The behaviour is shown in Fig. \ref{fig:fig7} (a), where the blue dotted line depicts the low Higgs oscillation with a big oscillation period, and the olive solid line is the high Higgs oscillation with a much smaller period.
The big period contrast of these two Higgs oscillations makes it quite easy to distinguish each other by naked eyes. In panels (b) and (c) of Fig. \ref{fig:fig7}, the corresponding results from the Fourier analysis are present and agree well with Volkov and Kogan's prediction in Eq.~\eqref{eq:Tana}. The low Higgs oscillation in the blue solid line on the left panel exhibits a clear low-frequency peak at $\omega\simeq1.31E_{F}$ (i.e., $T_{1}=4.80/E_{F}$), not far from the position of the high-frequency peak for the high Higgs oscillation marked by an arrow on the right panel. The lower peaks in (c) are from coupling effect of the Higgs oscillation to other excitations, and we have checked that this coupling can be weakened by taking a relatively larger driving amplitude $A$.
\begin{figure}[t]
\includegraphics[width=0.48\textwidth]{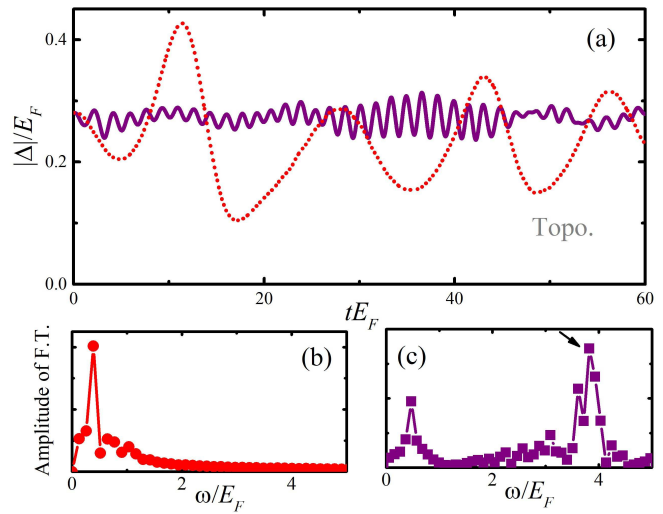}
\caption{(a) The oscillation of the order parameter $\left|\Delta(t)\right|$ after a periodic ramp in Eq.~\eqref{eq:hp} at $h_{i}=1.1E_{F}$. (b) and (c) are the corresponding Fourier analysis. Solid and dotted lines are the high and low Higgs oscillations, respectively.\label{fig:fig8}}
\end{figure}

In addition, we show the results in the topological phase in Fig. \ref{fig:fig8} with $h_{i}=1.1E_{F}$ where two Higgs oscillations are manifested in two significant oscillating behaviours of the order parameter. Different with the case in the BCS state, it is now much easier to distinguish two Higgs oscillations owing to their larger period contrast which may weaken the coupling between two Higgs oscillations following the resonance theory. The Fourier analysis in panels (b) and (c) of Fig. \ref{fig:fig8} further verifies the existence of two Higgs oscillations and shows an excellent agreement with the Volkov and Kogan's prediction in Eq.~\eqref{eq:Tana}.

In a word from Figs.~\ref{fig:fig7} and~\ref{fig:fig8}, we can clearly distinguish the high Higgs oscillation (solid lines) and the low Higgs oscillation (dotted lines) by this periodic ramp strategy in Eq.~\eqref{eq:hp} using their own driving frequency calculated from Volkov and Kogan's prediction at fixed $1.5$ periods. The oscillation amplitude of low Higgs oscillation is always larger than that of the high Higgs oscillation in both the BCS superfluid and the topological superfluid. We use this strategy to further verify the existence of two Higgs oscillations which can be determined by two energy branches in the quasi-particle spectrum. The large period contrast in the topological state makes it much easier to display these two Higgs oscillations than that in the BCS state.

\section{CONCLUSIONS} \label{sec:conclusion}

In summary, we theoretically probe and study two Higgs oscillations in a one-dimensional Raman-type spin-orbit-coupled Fermi superfluid, by solving the time-dependent BdG equations with three different ways to tune the effective Zeeman field. In contrast to the single Higgs oscillation in the conventional Fermi superfluid, we find two distinct Higgs oscillations in both the BCS and topological states when investigating numerically the oscillation of the order parameter. The Higgs oscillation can be well explained from the excitation in two quasi-particle energy spectrum, whose oscillation periods exhibit a great agreement with previous Volkov and Kogan's theoretical prediction over a large range of the Zeeman field except for crossing the phase transition point. Further research could be undertaken to thoroughly explore the oscillation behaviours related to other physical observables, such as density and spin polarization.

\begin{acknowledgments}
We are grateful for fruitful discussions with Shi-Guo Peng and kind help from Huaisong Zhao. This research was supported by the National Natural Science Foundation of China, Grants No. 11804177 (P.Z.), the Science Foundation of Zhejiang Sci-Tech University (ZSTU) No. 21062339-Y and China Postdoctoral Science Foundation No. 2020M680495 (X.L.C.).
\end{acknowledgments}

\bibliography{references}

\end{document}